\documentstyle[11pt,newpasp,twoside,epsf,psfig]{article}
\def\edcomment#1{\iffalse\marginpar{\raggedright\sl#1\/}\else\relax\fi}
\reversemarginpar

\begin{document}
\title{SB2 and eclipsing binaries with GAIA and RAVE}
 \author{Ulisse Munari}
 \affil{Osservatorio Astronomico di Padova INAF, sede di Asiago,\\ 36012 Asiago (VI), Italy}
 \author{Toma\v{z} Zwitter}
 \affil{University of Ljubljana, Department of Physics, Jadranska 19,\\ 1000 Ljubljana, Slovenia}
 \author{Eugine F. Milone}
 \affil{Rothney Astroph. Obs., Dept. of Physics and Astronomy, Univ. of Calgary, 2500 University Dr., NW
 Calgary, AB   T2N 1N4, Canada}

\begin{abstract}
The expected performance of GAIA satellite on eclipsing binaries is reviewed
on the basis of ($a$) combined Hipparcos and ground-based observations
mimicking GAIA data harvest, and ($b$) accurate simulations using the latest
instrument model. It is found that for a large majority of the 16\,000 SB2
eclipsing binaries that GAIA will discover 
at magnitudes $V\leq$13, the
orbital solutions and physical parameters will be derived with formal
accuracies better than 2\%. For the same stars the GAIA parallax errors will
be $\sim$5~{\em micro}-arcsec, i.e. an error of 0.5\% at 1 kpc, which will
allow iterative refine of the parameters and physics used in orbital
modeling. The detectability of SB2 binaries by the already up and running
spectral survey RAVE is discussed. It is found that all F-to-M SB2 binaries
showing a velocity separation $\geq$35~km~sec$^{-1}$ and a luminosity ratio
$\geq$0.5 will be recognized as such.
\end{abstract}

\section{GAIA}

GAIA is the ambitious flagship mission approved by ESA for a launch not
later than 2012, and possibly already by Q2 of 2010.  The mission
is designed to obtain extremely precise astrometry (in the {\sl
micro-}arcsec regime), multi-band photometry and medium/high resolution
spectroscopy for a large sample of stars.  The goals call for astrometry and
photometry (5 broad and 11 medium bands covering the optical range) to be
complete over the whole sky to $V=20$, and spectroscopy complete to
$V=17.5$. The spectroscopy is primarily meant to provide the radial velocities
and thus the 6$^{\rm th}$ component in the phase-space to complement the
other five provided by astrometry. Each star will be visited around a
hundred times during the five year mission life-time, in a fashion similar
to the highly successful {\sl Hipparcos} scanning mode. The spectra,
photometry and astrometry acquired at each transit {\em epoch} will be individually
accessible. GAIA has three telescopes/fields of view, roughly equally spaced
over a great circle on the sky, so that every $\sim$2 hours along
the 6 hours spin period, the same star will cross the field of view of
Astro-1 (where astrometric position and magnitude in 5 broad bands are
measured), of Astro-2 (identical to Astro-1), and of the Spectro-Photometric telescope
(where spectra and photometry in $\sim$11 medium bands are collected).

\begin{figure}[!t] 
\centerline{\psfig{file=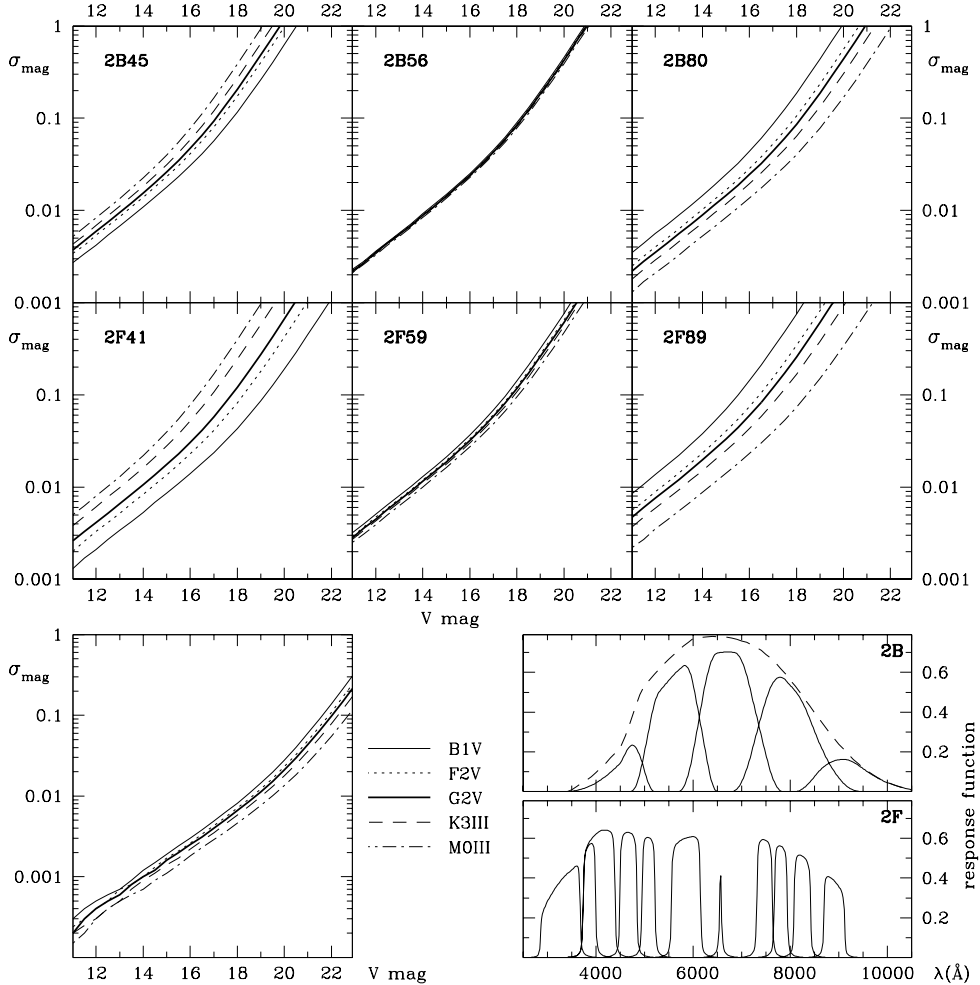,width=13.0cm}}
\caption{Errors of the magnitude measured over a single FOV transit for a
sample of broad (upper raw), medium (middle raw) and white-light (lower 
left) GAIA photometric bands. The current set
of broad and medium bands and the white-light photometric response of
astrometric CCDs are depicted at lower right (from data courtesy of C.Jordi, 
on behalf of GAIA Photometric Working group).}
\end{figure}

\begin{figure}[!t] 
\centerline{\psfig{file=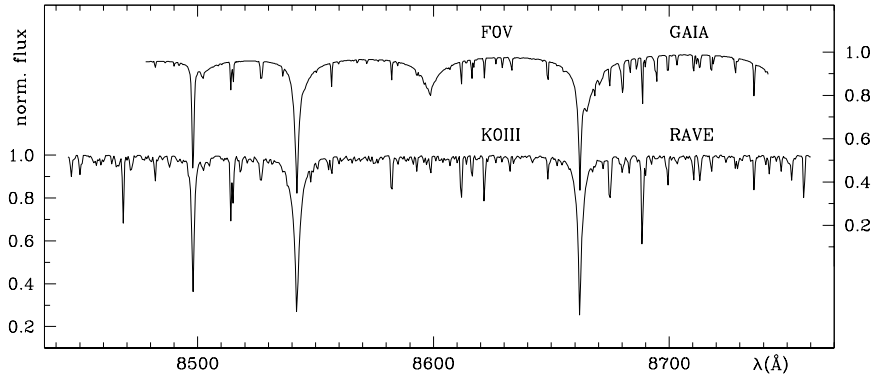,width=11.5cm}}
\caption{Comparison of GAIA (res. power 11\,500) and RAVE
(res. power 8\,000) spectra for two sample stars.}
\end{figure}

\begin{figure}[!b] 
\centerline{\psfig{file=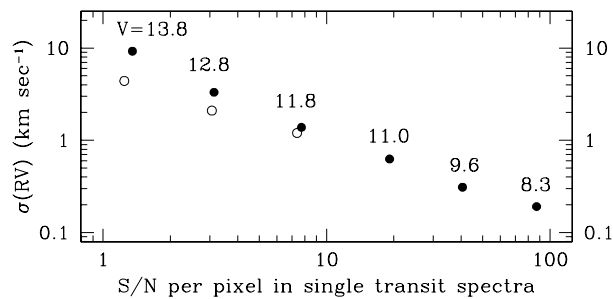,width=8.0cm}}
\caption{Errors of the GAIA radial velocities for single transit of single
G-K stars (dots from Munari et al. 2003, circles from Katz et al. 2004, to
be submitted).}
\end{figure}

\begin{figure}[!t] 
\centerline{\psfig{file=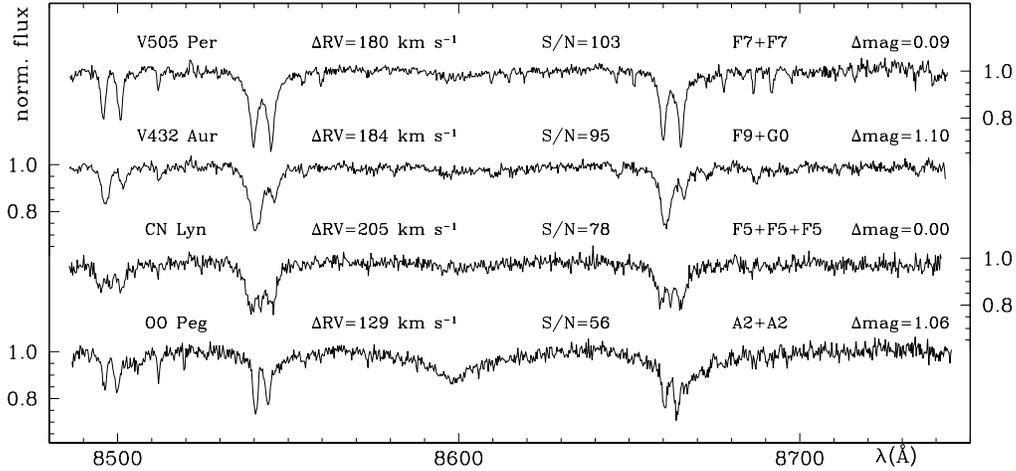,height=6.25cm,angle=270}}
\caption{Examples of GAIA-like ground-based spectra of SB2 eclipsing
binaries, where $\Delta$mag is the $I_{\rm C}$ mag difference of the two components
(from Munari et al. 2001, Zwitter et al. 2003, Marrese et al. 2004). Note the resolved 
triple system CN~Lyn.}
\end{figure}

\begin{figure}[!t] 
\centerline{\psfig{file=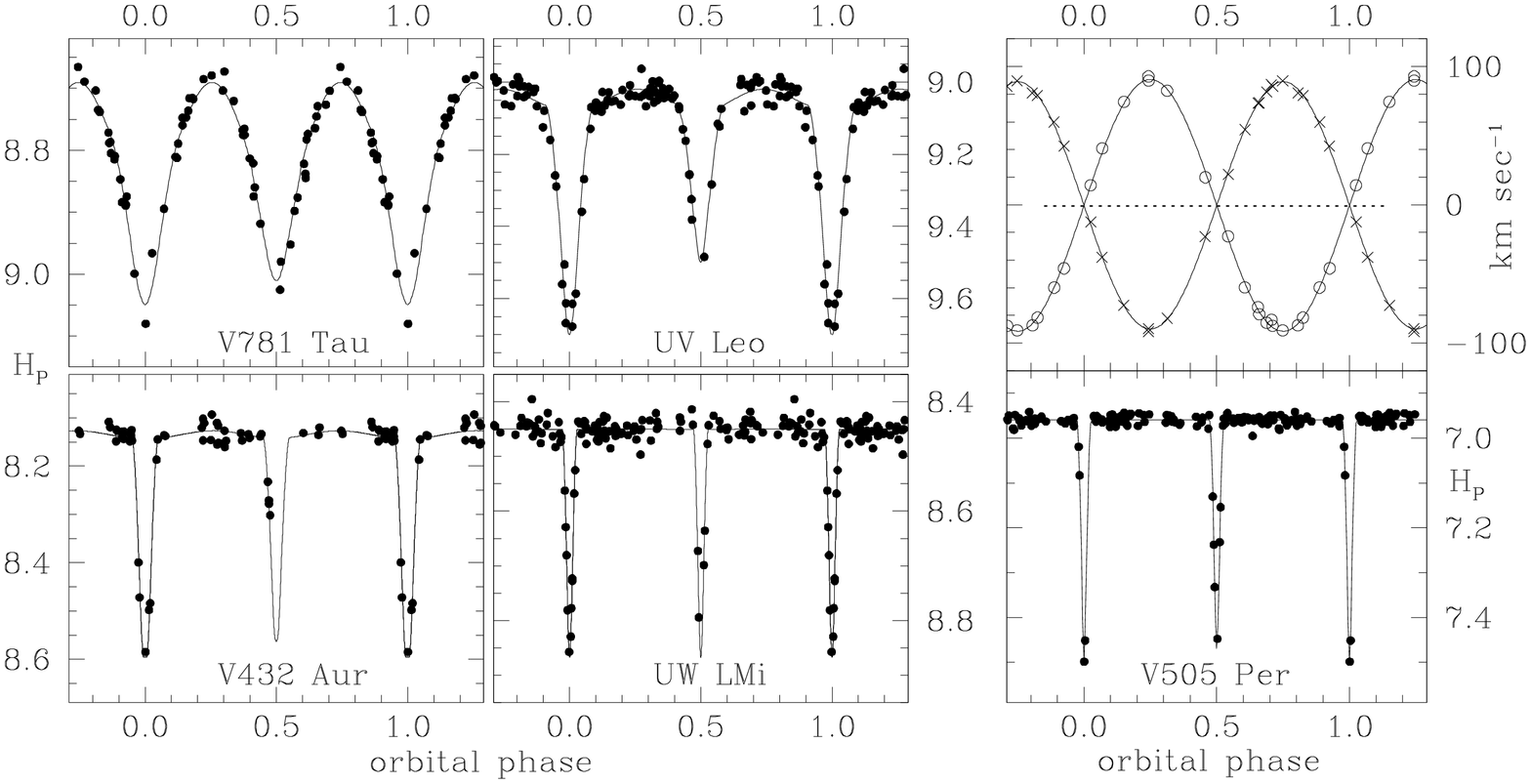,height=6.75 cm,angle=270}}
\caption{Hipparcos $H_P$ lightcurves of SB2 eclipsing binaries for a range of eclipse phase width
and number of points. The orbital solutions obtained in combination with radial velocities from 
ground-based GAIA-like spectroscopic observations are over-plotted (from Munari et al. 2001,
Zwitter et al. 2003, Marrese et al. 2004).}
\end{figure}

Hipparcos performed photometry in three bands. $H_P$, a sort of white light
covering the Johnson's $B$ and $V$ bands, has been the most accurate, with
individual measurements accurate to 0.01 mag for solar type stars brighter
than $V=7.5$.  The other two bands, $V_T$ and $B_T$, were significantly less
accurate. GAIA will carry instead a fully featured photometric system,
composed of a set of 5 broad and $\sim$11 medium bands, which are still
under fine optimization. Figure~1 illustrates the expected errors on the
magnitudes derived at each passage in the GAIA fields of view for three
sample broad bands, three sample medium bands and for the photometric
reading accompanying the {\em white light} astrometric measurement. Each
epoch measurement in any band and for any star brighter than $V=12$
(irrespective of the spectral type) will be accurate to better than
0.01~mag. In some bands (like in 2B56 or 2F59) the same limit is reached at
$V=14$. In white light, all individual measurements of stars brighter than
$V=18.5$ will be more accurate than 0.01 mag.

Field stars have colors corresponding to $\sim$G0 at $V=10$, $\sim$K0 at
$V=15$ and cooler spectral types at fainter magnitudes. Therefore,
optimization of spectroscopic observations has to be performed for G-K
stars. Low metallicity stars of the Halo
that covers a large fraction of the sky should be properly observed as well.
Wavelength region satisfying these constraints is that of the 
CaII triplet in the far red (cf. Figure~2). This  spectral region is 
interesting even for hotter stars, because it includes 
the head of the Paschen series and strong
multiplets \#1 and 8 of NI (for a detailed discussion see Munari 1999, 2002). 
Figure~3 illustrates the expected errors of GAIA radial velocities for
single field stars of G-K spectral types and for a single transit. All 
stars brighter than $V=14$ will have epoch radial velocities accurate to
10~km~sec$^{-1}$ or better, all those brighter than $V=11.6$ to
1~km~sec$^{-1}$ or better.

\begin{figure}[!t]
\centerline{\psfig{file=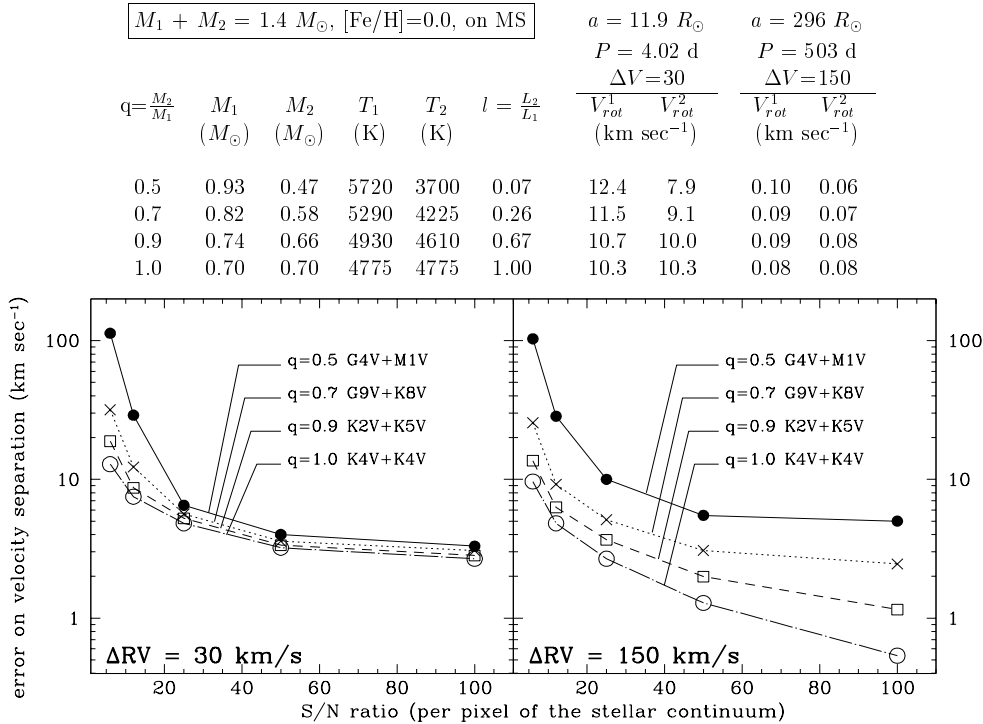,width=13.0cm}}
\caption{Results of Monte Carlo simulations to estimate the accuracy of GAIA
radial velocity measurements of SB2 binaries observed at maximum velocity
separation (orbital phases 0.25 or 0.75). Synthetic spectra of binaries with
a total mass of 1.4~M$_\odot$, solar metallicity and both components on the
Main Sequence were generated for the four given mass ratios. The latter fix
individual masses, temperatures, spectral types and luminosity ratios. Two
different orbital separations are assumed, so that eclipsing systems have
velocity separations of 30 and 150~km~sec$^{-1}$ at quadrature (this fixes
the synchronized rotational velocities).  Synthetic stellar spectra used as
input were calculated from Kurucz models using the $R=11\,500$ GAIA
resolving power. Radial velocity measurements were made with an IRAF
implementation of TODCOR routines. 30 and 150~km~sec$^{-1}$ corresponds to a
separation of 2.3 and 11.5 pixels in the focal plane of the GAIA
spectrograph, respectively. The simulations were performed for S/N=6, 12,
25, 50, 100 per pixel of the spectral continuum. The curves indicate the
2$\sigma$ error loci (i.e. 66\% of the trials gave errors smaller than
indicated). A $\Delta$rad.vel.=30~km~sec$^{-1}$=2.3~pixels corresponds
to the Rayleigh resolution criterion for intrinsically sharp lines, and the
accuracies displayed in the left panel go as expected with S/N and source
brightness contrast. The $\Delta$rad.vel.=150~km~sec$^{-1}$=11.3~pixels
on the right panel corresponds instead to the situation where the radial
velocity displacement is large enough to cause overlaps of unrelated
spectral lines of the two components. In this case the luminosity ratio is
the dominant factor governing the accuracy of the radial velocity
measurements.}
\end{figure}

\clearpage

\section{GAIA and the eclipsing binaries}

Previous reviews of SB2 eclipsing binaries as seen by GAIA that nicely complement
the present one have been given by Zwitter (2002, 2003), Zwitter \&\ 
Munari (2004) and Milone (2003).

Minimal velocity separation for a firm recovery of the SB2 nature for solar
type stars is of the order of 25-30~km~sec$^{-1}$. Scaling Figure~3
to binary star case we see that radial velocity of each star in an SB2 
binary with $V\leq13$ and $L_2/L_1 \geq$0.5 will be derived with an error 
not exceeding 10~km~sec$^{-1}$. With $\sim$100 spectra per star randomly distributed
in phase it may be expected that 1/3 of them will contribute 
to precise definition of semi-amplitudes $K_1$, $K_2$. If
their accuracy is better than 10~km~sec$^{-1}$, the resulting $K_1$ and
$K_2$ will be accurate to 1~km~sec$^{-1}$ or better. Figure~1 shows that at
$V=13$ the epoch photometry in any of the 5+11 bands is accurate to 0.01 mag
and in white light to an amazing 0.0006 mag for any of the considered 
stellar types. Therefore {\em all SB2 eclipsing binaries with $V\leq 13$ will be accurately observed
by GAIA}.

It is important to check if GAIA will collect enough photometric 
observations to properly constrain the shape of the light-curve, and
of the eclipse phases in particular.
The measurements in white light and broad bands are
essentially simultaneous and altogether $\sim$100 observations scattered over the five
year mission lifetime will be collected (the exact number depends on the ecliptic
coordinates and is generally comprised between 60 and 200). GAIA will also 
collect $\sim 170$ measurements in the medium bands (again nearly
simultaneous in all bands). Moreover, RVS sky mapper measurements 
obtained just before spectroscopic observations 
will provide additional $\sim$100 photometric points in a
further photometric band covering the 8480-8740~\AA\ spectroscopic
wavelength range. Therefore, combining the 2B80 broad band, the 2F89 medium
band and the RVS sky mappers, an average of about 370 independent photometric
measurements will be obtained at wavelengths close to the 
Cousins $I_{\rm C}$ band, and 270 measurements for all other bands.

Such a number
of photometric points should suffice for most of  the cases. 
We have embarked in a long term project to assess the accuracy of 
GAIA investigations of SB2 eclipsing binaries (Munari et al. 2001, Zwitter et
al. 2003, Marrese et al. 2004, Milone et al. 2004 to be submitted). 
Twelve systems have been studied so far, combining ground-based radial velocities acquired in the GAIA
wavelength range with Hipparcos $H_P$ photometry (to mimic GAIA photometric
mapping). Typical light (and radial velocity) curves are presented in
Figure~5, where orbital solutions are over-plotted. Such $H_P$ mapping is
generally sufficient to constrain the radii to 2-6\%, depending of the total
number of points and the width of the eclipse phases. With GAIA providing
3-- to 4--times more points, the fraction of SB2 eclipsing binary
systems for which the orbital solution will have formal errors in the 1-2\%
range should be remarkable. Similar conclusions were reached by Niarchos and 
Manimanis (2003) from their investigation in a GAIA-like mode of four over-contact systems.

Hipparcos derived a frequency of 0.8\% of eclipsing binaries among the stars
it surveyed. Applying this to star counts in the Galaxy, we may estimate 
that GAIA will discover 64\,000 eclipsing binaries with $V\leq 13$.
Applying Carquillat et al. (1982) conclusion that 1/4 of EBs are SB2, we end up with
about 16\,000 SB2 eclipsing binaries brighter than $V=13$ for which GAIA
should provide orbital solutions formally accurate to 1-2\%. This 
is a fantastic
number compared to $\leq$100 systems studied at similar accuracy by
ground-based observations so far. These 16\,000 stars will have their parallax
measured to 5 micro-arcsec accuracy which corresponds to an error 
$\leq$0.5\% in their distance if closer than 1~Kpc. 
Such extremely accurate parallaxes will be 
sed to directly constrain the values of several parameters 
(e.g.\ absolute luminosity) and physics used in orbital modeling, so that the atmospheric
structure and global parameters of eclipsing binaries could be refined to
{\em absolute} errors well below 1\%.

\begin{figure}[!t]
\centerline{\psfig{file=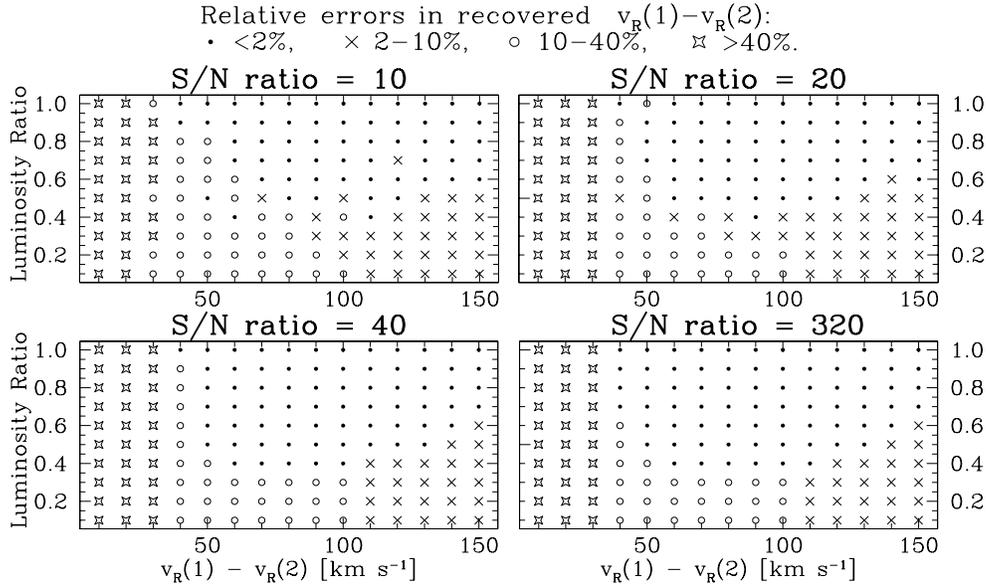,width=13.0cm}}
\caption{Results of Monte Carlo simulations to assess the recoverability of 
SB2 binaries in RAVE spectra. The difference between the true and recovered 
radial velocity difference of the two components is given for solar type
SB2 and a range of $\Delta$RV, S/N and luminosity ratio.}
\end{figure}
 
\section{RAVE and the SB2 binaries}
 
RAVE (RAdial Velocity Experiment) is an ambitious
program to conduct an all-sky survey (complete to $V$=15.5 mag) of 
radial velocities and metallicities of 50 million stars using the 1.2\,m
UK Schmidt Telescope of the Anglo-Australian Observatory (AAO), together
with a northern counterpart, over the period 2006 -- 2010. A key feature of
the RAVE project is a pilot program aiming to observe $10^5$ stars using the
existing 6dF facility in unscheduled bright time over the period 2003--2005.
This pilot program begun on April 11, 2003 and by the time this conference
was held it logged the observations of 20\,000 stars. The 6dF spectrograph
brackets the CaII triplet region adopted for GAIA (c. Figure~2). Spectra are
taken at a resolution of $R=8000$.

The target list of the pilot program includes a large fraction of the
$118\,000$ Hipparcos stars that are accessible from the southern hemisphere
as well as some of the $2\,539\,913$ stars of the Tycho-2 catalog. The
survey focuses on stars in the color range $0.4 < B-V < 0.8$.  For these
stars useful photometric parallaxes can be derived if meaningful
trigonometric parallaxes are not available.

The RAVE main survey should utilize a new Echidna-style multi fiber
spectrograph at the UK-Schmidt telescope and its northern counterpart. In
the UK-Schmidt planned version it consists of a 2250-spine fiber array
covering the full field of view (40 deg$^2$). 
The fiber positions are reconfigurable in $\approx 5$\,min.
A detailed description of RAVE and in particular of its science goals,
aiming to derive the kinematics and formation history of the Milky Way, can
be found in Steinmetz (2003).

RAVE observes any given target star only once, even if with remarkable S/N.
Therefore it is not possible to discover SB1 binaries or trace the motion of components in
SB2 systems. However, the respectable resolution and
high S/N of RAVE spectra allow it to detect all SB2 binaries with a suitable
luminosity ratio and velocity separation at the time of the observation.
Given the huge number of stars surveyed by RAVE, this would put useful
statistical constraints (on velocity separation and luminosity ratio as
a function of spectral type and metallicity) on the population of binary
stars in the Milky Way. Figure~7 presents the results of simulations carried
to out to infer the SB2 detectability by RAVE. They show that
all F--M SB2 binaries with a luminosity ratio $\geq$0.5 
and observed when velocity separation of their components is
$\geq$35~km~sec$^{-1}$ will be recognized
as such. And new SB2 binaries are already being discovered 
in the early data secured by RAVE in 2003!

\end{document}